\documentclass[aps,prc,twocolumn,showpacs,superscriptaddress]{revtex4}
\usepackage{graphicx}
\usepackage{dcolumn}
\usepackage{bm}

\begin{document}

\title{Measurements of transverse energy distributions in Au+Au collisions at 
$\sqrt{s_{NN}}= 200$ GeV}

\affiliation{Argonne National Laboratory, Argonne, Illinois 60439}
\affiliation{University of Bern, Bern, Switzerland CH-3012}
\affiliation{University of Birmingham, Birmingham, United Kingdom}
\affiliation{Brookhaven National Laboratory, Upton, New York 11973}
\affiliation{California Institute of Technology, Pasedena, California 91125}
\affiliation{University of California, Berkeley, California 94720}
\affiliation{University of California, Davis, California 95616}
\affiliation{University of California, Los Angeles, California 90095}
\affiliation{Carnegie Mellon University, Pittsburgh, Pennsylvania 15213}
\affiliation{Creighton University, Omaha, Nebraska 68178}
\affiliation{Nuclear Physics Institute AS CR, 250 68 \v{R}e\v{z}/Prague, Czech Republic}
\affiliation{Laboratory for High Energy (JINR), Dubna, Russia}
\affiliation{Particle Physics Laboratory (JINR), Dubna, Russia}
\affiliation{University of Frankfurt, Frankfurt, Germany}
\affiliation{Insitute  of Physics, Bhubaneswar 751005, India}
\affiliation{Indian Institute of Technology, Mumbai, India}
\affiliation{Indiana University, Bloomington, Indiana 47408}
\affiliation{Institut de Recherches Subatomiques, Strasbourg, France}
\affiliation{University of Jammu, Jammu 180001, India}
\affiliation{Kent State University, Kent, Ohio 44242}
\affiliation{Lawrence Berkeley National Laboratory, Berkeley, California 94720}
\affiliation{Massachusetts Institute of Technology, Cambridge, MA 02139-4307}
\affiliation{Max-Planck-Institut f\"ur Physik, Munich, Germany}
\affiliation{Michigan State University, East Lansing, Michigan 48824}
\affiliation{Moscow Engineering Physics Institute, Moscow Russia}
\affiliation{City College of New York, New York City, New York 10031}
\affiliation{NIKHEF, Amsterdam, The Netherlands}
\affiliation{Ohio State University, Columbus, Ohio 43210}
\affiliation{Panjab University, Chandigarh 160014, India}
\affiliation{Pennsylvania State University, University Park, Pennsylvania 16802}
\affiliation{Institute of High Energy Physics, Protvino, Russia}
\affiliation{Purdue University, West Lafayette, Indiana 47907}
\affiliation{University of Rajasthan, Jaipur 302004, India}
\affiliation{Rice University, Houston, Texas 77251}
\affiliation{Universidade de Sao Paulo, Sao Paulo, Brazil}
\affiliation{University of Science \& Technology of China, Anhui 230027, China}
\affiliation{Shanghai Institute of Applied Physics, Shanghai 201800, P.R. China}
\affiliation{SUBATECH, Nantes, France}
\affiliation{Texas A\&M University, College Station, Texas 77843}
\affiliation{University of Texas, Austin, Texas 78712}
\affiliation{Tsinghua University, Beijing, P.R. China}
\affiliation{Valparaiso University, Valparaiso, Indiana 46383}
\affiliation{Variable Energy Cyclotron Centre, Kolkata 700064, India}
\affiliation{Warsaw University of Technology, Warsaw, Poland}
\affiliation{University of Washington, Seattle, Washington 98195}
\affiliation{Wayne State University, Detroit, Michigan 48201}
\affiliation{Institute of Particle Physics, CCNU (HZNU), Wuhan, 430079 China}
\affiliation{Yale University, New Haven, Connecticut 06520}
\affiliation{University of Zagreb, Zagreb, HR-10002, Croatia}

\author{J.~Adams}\affiliation{University of Birmingham, Birmingham, United Kingdom}
\author{M.M.~Aggarwal}\affiliation{Panjab University, Chandigarh 160014, India}
\author{Z.~Ahammed}\affiliation{Variable Energy Cyclotron Centre, Kolkata 700064, India}
\author{J.~Amonett}\affiliation{Kent State University, Kent, Ohio 44242}
\author{B.D.~Anderson}\affiliation{Kent State University, Kent, Ohio 44242}
\author{D.~Arkhipkin}\affiliation{Particle Physics Laboratory (JINR), Dubna, Russia}
\author{G.S.~Averichev}\affiliation{Laboratory for High Energy (JINR), Dubna, Russia}
\author{Y.~Bai}\affiliation{NIKHEF, Amsterdam, The Netherlands}
\author{J.~Balewski}\affiliation{Indiana University, Bloomington, Indiana 47408}
\author{O.~Barannikova}\affiliation{Purdue University, West Lafayette, Indiana 47907}
% ...

\collaboration{STAR Collaboration}\noaffiliation

\date{July 02, 2004.}

\begin{abstract}

Transverse energy ($E_T$) distributions have been measured for Au+Au collisions at 
$\sqrt{s_{NN}}= 200$ GeV by the STAR collaboration at RHIC. $E_T$ is constructed from its 
hadronic and electromagnetic components, which have been measured separately. $E_T$ 
production for the most central collisions is well described by several theoretical models 
whose common feature is large energy density achieved early in the fireball evolution. The 
magnitude and centrality dependence of $E_T$ per charged particle agrees well with 
measurements at lower collision energy, indicating that  the growth in $E_T$ for larger 
collision energy results from the growth in particle production. The electromagnetic 
fraction of the total $E_T$ is consistent with a final state dominated by mesons and 
independent of centrality.

\end{abstract}

\maketitle

\section{Introduction}

\def\ET{\mbox{$E_T$}}

\def\hadET{\mbox{$E_T^{had}$}}
\def\emET{\mbox{$E_T^{em}$}}

\def\dETdeta{$dE_T/d\eta$}
\def\avdETdeta{$\langle dE_T/d\eta \rangle$}
\def\dNchdeta{$dN_{ch}/d\eta$}
\def\avdNchdeta{$\langle dN_{ch}/d\eta \rangle$}

\def\dETdy{\mbox{$dE_T/dy$}}

\def\pT{\mbox{$p_T$}}

\def\Npart{\mbox{$N_{part}$}}

High energy nuclear collisions at the Relativistic Heavy Ion Collider (RHIC) \cite{refRHIC} 
have opened a new domain in the exploration of strongly interacting matter at very high 
energy density. High temperatures and densities may be generated in the most central
(head-on) nuclear collisions, perhaps creating the conditions in which a phase of 
deconfined quarks and gluons exists \cite{refBlaizot,refJacbWang}. The fireball produced in 
such collisions undergoes a complex dynamical evolution, and understanding of the 
conditions at the hot, dense early phase of the collision requires understanding of the 
full reaction dynamics.

Transverse energy \ET\ is generated by the initial scattering of the partonic constituents 
of the incoming nuclei and possibly also by reinteractions among the produced partons and 
hadrons \cite{refJacob,refWang}. If the fireball of produced quanta breaks apart quickly 
without significant reinteraction, the observed transverse energy per unit rapidity \dETdy\ 
will be the same as that generated by the initial scatterings. At the other extreme, if the
system reinteracts strongly, achieving local equilibrium early and maintaining it 
throughout the expansion, \dETdy\ will decrease significantly during the fireball evolution 
due to the longitudinal work performed by the hydrodynamic pressure
\cite{refMatsui,refEskola}. This decrease will however be moderated by the buildup of 
transverse hydrodynamic flow, which increases \ET\ \cite{refKolb}. Finally, gluon 
saturation in the wave function of the colliding heavy nuclei can delay the onset of 
hydrodynamic flow, reducing the effective pressure and thereby also reducing the difference 
between initially generated and observed \ET\ \cite{refDumitru}.

\ET\ production in nuclear collisions has been studied at lower $\sqrt{s}$ at the AGS and 
CERN \cite{refYoung, refWA80, refBarrete2, refNA49, refWA98} and at RHIC \cite{refPHENIX}. 
Within the framework of boost-invariant hydrodynamics \cite{refBjorken},
these measurements suggest that energy densities have been achieved at the SPS 
\cite{refNA49} that exceed the deconfinement energy density predicted by Lattice QCD 
\cite{refKarsch}. However, from the foregoing discussion it is seen that several competing 
dynamical effects can contribute to the observed \dETdy. While the measurement of \ET\ 
alone cannot disentangle these effects, a systematic study of \ET\ together with other 
global event properties, in particular charged multiplicity and mean transverse momentum
$\left\langle p_T\right\rangle$, may impose significant constraints on the collision 
dynamics \cite{refKolb}.

In this paper we report the measurement of \ET\ distributions from Au+Au collisions at 
$\sqrt{s_{NN}}=200$ GeV per nucleon-nucleon pair, measured by the STAR detector at RHIC. 
\ET\ is measured using a patch of the STAR Electromagentic Calorimeter, with acceptance 
$0<\eta<1$ and $\Delta\phi=60^{\circ}$, together with the STAR Time Projection Chamber.  
\ET\ is separated into its hadronic and electromagnetic components, with the latter 
dominated by $\pi^0$ and $\eta$ decays. The centrality dependence of \ET\ and \ET\ per 
charged particle are studied, and comparisons are made to models and to measurements at 
lower energy.

A high temperature deconfined phase could be a significant source of low to intermediate 
\pT\ photons \cite{refThermalPhotons}. An excess of photons above those expected from 
hadronic decays has been observed at the SPS for $p_T>1.5$ GeV/c \cite{refWA98eta}. We 
investigate this effect through the study of the electromagnetic component of \ET.

Section II describes the experimental setup used for the analysis. Section III presents the 
analysis of the hadronic component of the transverse energy. In section IV, the analysis of 
the electromagnetic transverse energy is presented. In section V we discuss the scaling of 
\ET\ with the energy of the colliding system and the number of participants \Npart\ and 
binary collisions $N_{bin}$ \cite{refNpartNbin}, together with theoretical expectations for 
this scaling. We also discuss the behavior of the electromagnetic component of the 
transverse energy with the collision energy and centrality. Section VI is a summary and 
discussion of the main results.

\section{STAR experiment}

This analysis is based on 150K minimum bias Au+Au collisions measured by the STAR detector 
in the 2001 RHIC run. STAR \cite{refSTAR} is a large acceptance, multi purpose experiment 
comprising several detector systems inside a large solenoidal magnet. In the following, we 
describe the detectors which are relevant to the present analysis.

The barrel Electromagnetic Calorimeter (EMC) \cite{refEMC} is a lead-scintillator sampling 
electromagnetic calorimeter with equal volumes of lead and scintillator. It has a radius of 
2.3 m and is situated just inside the coils of the STAR solenoidal magnet.  The 
electromagnetic energy resolution of the detector is $\delta E/E \sim 
16\%/\sqrt{E(\textrm{GeV})}$.  The results presented in this work used the first EMC patch 
installed for the 2001 RHIC run consisting of 12 modules, $\sim 10$\% of the full planned 
detector, with coverage $0 < \eta < 1$ and $\Delta\phi=60^{\circ}$. Each EMC module is 
composed of 40 towers (20 towers in $\eta$ by 2 towers in $\phi$) constructed to project to 
the center of the STAR detector. The transverse dimensions of a tower are approximately 10 
x 10 cm$^2$, which at the radius of the front face of the detector correspond to a phase 
space interval of 
($\Delta\eta,\Delta\phi$)=(0.05, 0.05). The tower depth is 21 radiation lengths ($X_0$), 
corresponding to approximately one hadronic interaction length. When fully installed, the 
complete barrel will consist of 120 modules with pseudorapidity coverage $-1 < \eta < 1$ 
and full azimuthal coverage.

The Time Projection Chamber (TPC) \cite{refTPC} has a pseudorapidity coverage of $|\eta| < 
1.2$ for collisions in the center of STAR, with full azimuthal coverage. In this work, the 
acceptance of the measurement was limited by the acceptance of the EMC. For charged tracks 
in the acceptance, the TPC provides up to 45 independent spatial and specific ionization 
$dE/dx$  measurements. The $dE/dx$ measurement in combination with the momentum measurement 
determines the particle mass within limited kinematic regions.

The magnetic field was 0.5 Tesla. TPC track quality cuts included $z$-coordinate 
(longitudinal axis) selection of the collision vertex within 20 cm of the TPC center and a 
minimum TPC track space point cut of 10. Typical TPC momentum resolution for the data in 
this work is characterized by $\delta k/k \sim 0.0078+0.0098 \cdot p_T$(GeV/$c$) 
\cite{refTPC} in which $k$ is the track curvature, proportional to $1/p_T$. Typical 
resolution of $dE/dx$ measurement is $\sim 8$\%.  Additional discussion of TPC analysis is 
given in the following sections and a more detailed description of the TPC itself can be 
found in Ref. \cite{refTPC}.

The event trigger consisted of the coincidence of signals from the two Zero Degree 
Calorimeters (ZDC) \cite{refZDC}, located at $\theta<2$ mrad about the beam downstream of 
the first accelerator dipole magnet and sensitive to spectator neutrons. These calorimeters 
provide a minimum bias trigger which, after collision vertex reconstruction, corresponds to 
$97 \pm 3\%$ of the geometric cross section $\sigma_{geom}^{Au+Au}$. The events were 
analysed in centrality bins based on the charged particle multiplicity in $|\eta |<0.5$.

The procedures used in the analysis provide independent measurement of  electromagnetic 
transverse energy and the transverse energy carried by charged hadrons. This latter 
quantity, corrected to take into account the contribution of the long-lived neutral 
hadrons, is designated the hadronic transverse energy. The hadronic component of the 
transverse energy is obtained from momentum analyzed tracks in the TPC while the 
electromagnetic fraction is derived from the Electromagnetic Calorimeter data corrected for 
hadronic contamination using TPC tracking. In the following sections we describe how each 
of these contributions were analyzed to obtain the total transverse energy \ET\ 
measurement.

\section{Hadronic Transverse Energy ($E_T^{had}$)}

The hadronic transverse energy \hadET\ is defined as

\begin{equation}
\label{eqHadEt1}
  E_T^{had}=\sum_{hadrons} E^{had} \sin\theta
\end{equation}

\noindent
where the sum runs over all hadrons produced in the collision, except $\pi^0$ and $\eta$.  
$\theta$ is the polar angle relative to the beam axis and the collision vertex position. 
$E^{had}$ is defined for nucleons as kinetic energy, for antinucleons as kinetic energy 
plus twice the rest mass, and for all other particles as the total energy. \hadET\ is 
measured using charged particle tracks in the TPC via:

\begin{equation}
\label{eqHadEt2}
  E_T^{had}=C_{0} \sum_{tracks} C_1(ID,p) E_{track}(ID,p) \sin\theta
\end{equation}
\noindent
The sum includes all tracks from the primary vertex in the ranges $0 < \eta < 1$ and 
$\Delta\phi=60^\circ$. $C_{0}$ is a correction factor defined as:

 \begin{equation}
\label{CZero}
  C_0 =\frac{1}{f_{acc}} \frac{1}{f_{p_Tcut}} \frac{1}{f_{neutral}}
\end{equation}

\noindent
that includes the effective acceptance $f_{acc} = \Delta\phi/2\pi$, the correction 
$f_{neutral}$, for long lived neutral hadrons not measured by the TPC, and $f_{p_Tcut}$, 
for the TPC low momentum cutoff. $E_{track}(ID,p)$ is the energy associated with the 
particular track, either total or kinetic, as described above, computed from the measured 
momentum and particle identity (ID) as described below. The factor $C_{1}(ID,p)$ is defined 
as:

\begin{equation}
\label{COne}
  C_1(ID,p) ={f_{bg}(p_T)}\frac {1}{f_{notID}}\frac{1}{\mathrm{eff}(p_T)}
\end{equation}

\noindent
which includes the corrections for the uncertainty in the particle ID determination, 
$f_{notID}$, momentum dependent tracking efficiency, $\mathrm{eff}(p_T)$, and momentum 
dependent backgrounds, $f_{bg}(p_T)$. Next, we describe the corrections included in these 
two factors. 

Particle identification was carried out using the measurements of momentum and truncated 
mean specific ionization $\langle dE/dx \rangle$ in the TPC. For $p_T< 1$ GeV/$c$, 
assignment was made to the most probable particle type relative to the Bethe-Bloch 
expectation. Particles were assumed to be pions if $\langle dE/dx \rangle$ differed from 
this expectation by more than three standard deviations, or if $p_T > 1$ GeV/$c$. The 
uncertainty in this procedure was gauged by calculating \hadET\ for $p_T < 1$ GeV/$c$ both 
with the correct particle assignments and with all particles assumed to be pions. The ratio 
of these values for \hadET\ is applied as a correction for particles that can not be 
identified, yielding an overall correction factor to \hadET\ of $f_{notID} = 0.96 \pm 
0.02$. Because this correction was calculated from low momentum particles it does not 
account for the centrality variations in the particle ratios with $p_T > 1$ GeV/c 
\cite{refParticlePhenix}. On the other hand, particles at $p_T > 1$ GeV/c account for about 
20\% of the total number of particles. Taking into account the centrality-dependence 
increases in the p/$\pi$ and K/$\pi$ ratios at higher \pT\ generates a change in the 
estimated hadronic \ET\ on the order of 2\%, which is within the systematic error of 
$f_{notID}$.

Only tracks with a transverse momentum $p_T > 0.15$ GeV/$c$ were accepted because the 
tracking efficiency drops rapidly below this value. GEANT \cite{refGEANT} detector 
simulations of HIJING \cite{refHIJING} events demonstrate that this cut excludes 5\% of the 
total \hadET. A correction $f_{p_Tcut}$ for this effect is included in $C_{0}$. Taking all 
simulated tracks for $p_T < 0.15$ GeV/$c$ and calculating the energy assuming pions in two 
extreme cases, one with momentum $p = 0$ and other with $p = 0.15$ GeV/$c$ resulted in a 
variation of 3\% in \hadET, which was assigned as the systematic uncertainty due to this 
correction.

Since only primary charged tracks measured by the TPC are used in this analysis, we need to 
correct \hadET\ to include the contribution from long-lived neutral hadrons, principally 
$n$($\bar{n}$), $K^0_L$, $K^0_S$ and $\Lambda$($\bar{\Lambda}$). The correction factor 
applied to the data, defined as $f_{neutral}=E_T^{charged}/(E_T^{charged}+E_T^{neutral})$, 
can be estimated using measurements by STAR at 130 GeV \cite{refSTARpBar, refSTARpBarpErr, 
refSTARNegChHad, refSTARKaon, refSTARLambda}. We assume, based on HIJING simulations, that 
$f_{neutral}$ does not change significantly from 130 GeV to 200 GeV. We assume that the 
spectrum shape and yield for $K^0_L$ are the same as for $K^0_S$. The same approximation 
was applied in the case of $n$($\bar{n}$), after subtraction of the contribution from 
$\bar\Lambda$ decays from the measured $\bar{p}$ yield, and the measured STAR $\bar{p}/p$ 
ratio \cite{refSTARpBar}. Using this procedure we obtained a value of $f_{neutral}= 0.81 
\pm 0.02$. The uncertainty on this correction was estimated from the uncertainties in the 
measured STAR spectra. A cross check of these correction factors utilizing 200 GeV 
measurements \cite{refSTAR200GeV} generates variations well within the assigned systematic 
uncertainties.

The correction $f_{bg}(p_T)$ for background, consisting of electrons, weak-decays and 
secondary tracks that are misidentified as primary, depends on the type of the track and is 
divided into two separate corrections. The first is for the electrons which are 
misidentified as hadrons. This correction was estimated using the shape of the electron 
spectrum obtained from HIJING and GEANT simulations and the absolute yield from STAR data 
in the region where electrons are identified with high purity using the TPC $dE/dx$ 
measurements (essentially below 300 MeV). The second term is due to weak decays, which have 
been included in $f_{neutral}$ and therefore must be excluded from the primary track 
population to avoid double counting of their energy. In this case, the correction factor 
was calculated by embedding simulated particles into real events. By comparison between the 
simulated particles and the reconstructed ones, the fraction of secondary tracks assigned 
as primary was evaluated. $\Lambda$ and $K^0$ were simulated using the experimental yield 
and spectral shape measured by STAR \cite{refSTARLambda, refSTARKaon}.

The TPC reconstruction efficiency, $\mathrm{eff}(p_T)$, was also determined by embedding 
simulated tracks into real events and comparing the simulated input and the final 
reconstructed event. In order to evaluate the effect of different particle species in the 
reconstruction efficiency, pions, kaons and protons were embedded in the real events. In 
this work, the charged track efficiency correction is the average, weighted by the relative 
populations of each of these species. The track reconstruction efficiency depends on the 
transverse momentum of the tracks and the total track density. For central events the 
efficiency  is about 0.7 for tracks with $p_T = 0.25$ GeV/$c$ and reaches a plateau at 
about 0.8 for $p_T > 0.4$ GeV/$c$. This efficiency correction includes the efficiency for 
track reconstruction, the probability for track splitting, ghost tracks and dead regions of 
the TPC. 

The resulting systematic uncertainties, taking into account all corrections, combine in 
quadrature to a systematic uncertainty estimate of 6.1\% on \hadET. In Table 
\ref{tableHadCor} we summarize all individual corrections and the corresponding systematic 
uncertainties.

\begin{table} [h]
 \caption{Corrections and systematic uncertainties for hadronic energy \hadET\ for the 5\% 
most central collisions.  The quadrature sum of all the systematic uncertainties results in 
a total of 6.1\%. The upper part of the table shows the global corrections included in 
$C_0$ and the bottom part shows track wise corrections included in $C_1(ID,p)$. In this 
case, the correction values for $p_T=0.25$ GeV/$c$ and $1.0$ GeV/$c$ are shown.}

\begin{ruledtabular}
 \begin{tabular}{cc}
 Description & Correction \\
 \hline
 $f_{p_Tcut}$           & 0.95 $\pm$ 0.03 \\
 $f_{neutral}$          & 0.81 $\pm$ 0.02 \\
 \hline
 $f_{notID}$            & 0.96 $\pm$ 0.02 \\
 $f_{bg}(p_T)$          & 0.84 $\pm$ 0.02 (0.25 GeV/$c$) \\
                        & 0.94 $\pm$ 0.02 (1.0 GeV/$c$) \\
 $\mathrm{eff}(p_T)$    & 0.70 $\pm$ 0.04  (0.25 GeV/$c$) \\
                        & 0.80 $\pm$ 0.04  (1.0 GeV/$c$) \\ 
 \end{tabular}

 \end{ruledtabular}
\label{tableHadCor}
\end{table}

HIJING and GEANT simulations of \hadET\ measured in the acceptance of this study generate 
event-wise fluctuations of about 10\%. Simulations utilizing a substantially larger 
acceptance ($0<\eta<1$, $0<\phi<2\pi$) generate event-wise fluctuations of about 4\%, with 
this latter resolution resulting mainly from tracking efficiency and neutral hadron 
corrections.

The final \hadET\ distribution is corrected for vertex reconstruction efficiency. 
Peripheral events have lower vertex reconstruction efficiency which suppresses the the 
transverse energy distributions with respect to more central events. The vertex 
reconstruction efficiency depends on the number of tracks measured in the TPC and varies 
from 70\% to 97\%.

\section{Electromagnetic Transverse Energy ($E_T^{em}$)}

The electromagnetic transverse energy \emET\ is the sum of the measured transverse energy 
of electrons, positrons and photons. The largest fraction of this energy comes from $\pi^0$ 
decays. Electrons (and positrons) are included because more than 90\% of them are produced 
in the conversion of photons in detector materials. The energy of photons and electrons is 
fully measured by the calorimeter. There is also a contribution from charged and neutral 
hadrons produced in the collision that is significant and must be subtracted to permit a 
measurement of \emET. In order to remove the hadronic contribution from the measurement, we 
studied the full spatial profiles of energy deposition by identified hadrons in the EMC. An 
extensive experimental library of hadronic shower clusters in the calorimeter has been 
obtained which, in conjunction with TPC tracking, allow a correction for the hadronic 
background in the calorimeter.  

Section IV.A discusses the calibration of the EMC using minimum ionizing particles and 
electrons, while section IV.B discusses the correction for hadronic energy deposition in 
the EMC and section IV.C discusses the determination of \emET.

\subsection{Calibration of EMC}

Hadrons striking the EMC deposit a widely fluctuating fraction of their incident energy 
through hadronic showers. In addition, $\sim 30-40$\% of all high energy charged hadrons 
penetrate the entire depth of the EMC without hadronic interaction. If such a non-showering 
primary charged hadron has sufficient momentum, it will behave like a minimum ionizing 
particle (MIP) as it transits each of the scintillator layers, resulting in uniform total 
energy deposition which will be nearly independent of the incident momentum but will vary 
linearly with the total thickness of scintillator traversed. Due to the projective nature 
of the detector the total length of the scintillator increases with increasing $\eta$. The 
MIP peak therefore varies from 250 MeV at small $\eta$ to 350 MeV at large $\eta$. The 
absolute energy of the MIP peak and its $\eta$ dependence was determined from cosmic rays 
and test beam measurements \cite{refEMCcalib}.

\begin{figure}[t]
 \includegraphics[width=0.45\textwidth]{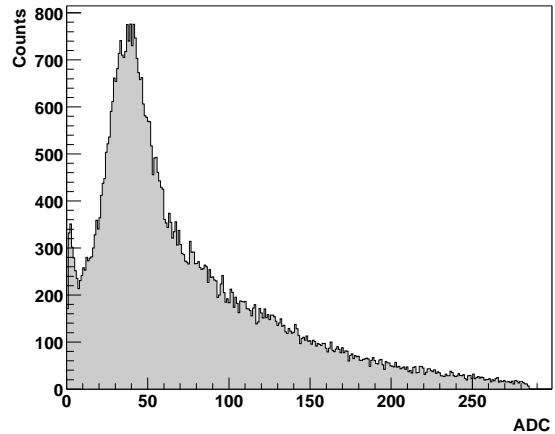}
 \caption{Typical MIP spectrum. The hits correspond to isolated  tracks with $p > 1.25$ 
GeV/$c$ which project to EMC towers. The  peak corresponds to the energy deposited by 
non-showering hadrons  (MIP peak).}
 \label{figMIP}
\end{figure}

The use of MIP particles to calibrate the EMC in situ is convenient and provides a 
precision tool to track the calibration of the detector over time. In a procedure to 
minimize systematic uncertainties in the calibration, tracks with $p > 1.25$ GeV/$c$ in the 
TPC from relatively low multiplicity events are extrapolated to the EMC towers where they 
are required to be isolated from neighboring charged tracks in a 3x3 tower patch 
($\Delta\eta$ x $\Delta\phi = 0.15$ x 0.15) which has a minimum size of $\sim$ 30 cm x 30 
cm ($\eta=0$). Fig. \ref{figMIP} shows a typical MIP spectrum measured under these 
conditions using minimum bias Au+Au events. This example shows the pseudorapidity interval 
$0.2 < \eta < 0.3$. Similar spectra are observed in all $\eta$ bins and provide an absolute 
calibration in the energy range less than $\sim 2$ GeV, with an estimated systematic 
uncertainty of $\sim 5$\% \cite{refEMCcalib}. 

An absolute calibration over a much wider energy range is obtained using identified 
electrons tracked with the TPC. This was done by selecting high momentum ($1.5<p<5.0$ 
GeV/$c$) electrons reconstructed in the TPC. Electron candidates are selected by $dE/dx$ 
measurement in the TPC. Although the purity of the electron candidates sample in this 
momentum range is poorer than for low momentum, the hadronic rejection factor obtained from 
the TPC $dE/dx$ provides a clear electron signal in the calorimeter. Bethe-Bloch 
predictions for $dE/dx$ of electrons and heavy particles show that the main background in 
this momentum range comes from deuterons and heavier particles as well as the tails of the 
distributions of protons and lower mass particles. In order to minimize systematic 
uncertainties in this procedure, only tracks having number of space points greater than 25 
were used, as such ``long tracks" exhibit better $dE/dx$ resolution. It was also required 
that the track should be isolated in a 3x3 tower patch in the calorimeter.

\begin{figure}[t]
 \includegraphics[width=0.45\textwidth]{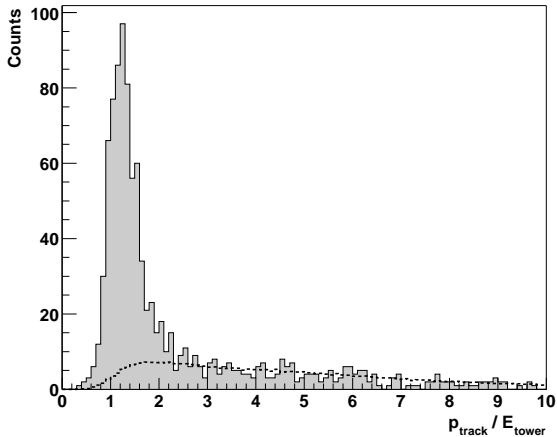}
 \caption{$p/E_{tower}$ spectrum for electron candidates, selected  through $dE/dx$ from 
the TPC, with $1.5 < p < 5.0$ GeV/$c$. A well  defined electron peak is observed. The 
dashed line corresponds  to the hadronic background in the $dE/dx$-identified electron  
sample.}
 \label{figElec}
\end{figure}

\begin{figure}[t]
 \includegraphics[width=0.45\textwidth]{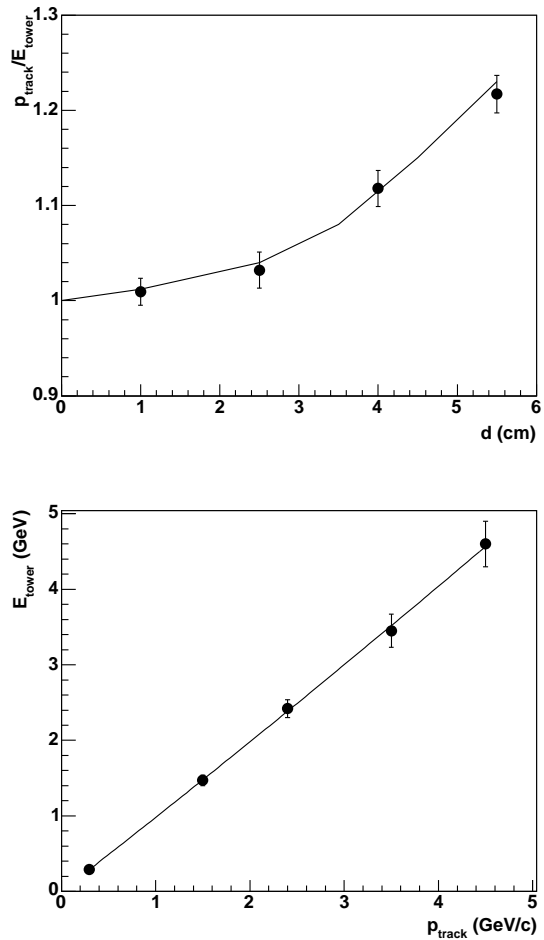}
 \caption{Upper plot: points are measured $p/E_{tower}$ electron peak position as a 
function of the distance to the center of the tower. The solid line is from a calculation 
based on a full GEANT simulation of the detector response to electrons. Lower plot: points 
show measured energy deposited by electrons in the tower as a function of the momentum for 
distances to the center of the tower smaller than 2.0 cm. The first point is the electron 
equivalent energy of the minimum ionizing particles. The solid line is a second order 
polynomial fit of the data.}
 \label{figElec2}
\end{figure}

As the final electron identifier, the energy, $E_{tower}$, deposited in the tower hit by 
the track is compared to the momentum, $p$, of the track in the range $1.5 < p < 5.0$ 
GeV/$c$. Fig. \ref{figElec} shows the $p/E_{tower}$ spectrum for the electron candidates in 
which it is possible to see a well defined electron peak.  The residual hadronic background 
in this figure can be evaluated by shifting the $dE/dx$ selection window toward the pion 
region. The resulting estimate of the hadronic background is shown as a dashed line in the 
figure. After hadronic background subtraction, the peak position is still not centered at 1 
due to the energy leakage to neighboring towers that is not taken into account in this 
procedure. The amount of leakage depends on the distance to the center of the tower hit by 
the electron and will shift the peak position to higher values as this distance increases. 
As shown in Fig. \ref{figElec2}, this effect is reproduced well by the full GEANT 
simulations of the detector response when it is hit by electrons in the momentum range used 
in this calibration procedure. The upper plot of Fig. \ref{figElec2} shows the position of 
the electron $p/E_{tower}$ peak as a function of this distance. The solid line is a 
prediction from GEANT simulations. The measurements are in  good agreement with the 
simulations. Fig. \ref{figElec2} lower plot shows the energy deposited in the calorimeter 
tower as a function of its momentum for electrons in the case where the distance to the 
center of the tower is smaller than 2.0 cm (A distance of 5 cm corresponds to the border of 
a tower at $\eta=0$. The border of the tower at $\eta=1$ is located 7.5 cm from its 
center). The first point is the electron equivalent energy of the minimum ionizing 
particles. A fit to the data using the second order polynomial of type $f(x) = a_0+ a_1x + 
a_2x^2$ is represented by the solid line. The coefficients are $a_0 = 0.01 \pm 0.08$ GeV, 
$a_1 = 0.98 \pm 0.11$ $c$ and $a_2 = 0.01 \pm 0.03$ (GeV/$c^2$)$^{-1}$. The values of $a_0$ 
and $a_2$ are consistent with zero within errors. The small magnitude of these errors 
indicates that the detector response to electrons is very linear up to $p=5$ GeV/$c$. 

By combining the MIP calibration and the electron calibration of the EMC we obtain an 
overall estimated systematic uncertainty of less than 2\% on the total energy measured by 
the calorimeter. The stability of the detector response was evaluated by monitoring the 
time dependence of the shape of the raw ADC spectra for each tower, which is the tower 
response for all particles that reach the calorimeter. In order to have enough statistics, 
each time interval is larger than one day of data taking but smaller than 2 days, depending 
on the beam intensity during that period. The overall gain variation of the detector was 
less than 5\% for the entire RHIC run. The results reported in this paper were obtained 
from 3 consecutive days of data taking, in order to minimize any uncorrected effect due to 
gain variations in the detector.

\subsection{Energy Deposited by Hadrons in the EMC}

As discussed above, for the purposes of measuring electromagnetic energy production it is 
essential to subtract the hadronic energy deposition in the calorimeter.  For charged 
hadrons, the hit locations on the calorimeter are well determined and if isolated, a 
cluster of energy is readily identified.  In the dense environment of Au+Au collisions, 
however, it is difficult to uniquely identify the energy deposition associated with a 
specific hadron track. In this limit, which is relevant for the present measurement, we 
subtract an average energy deposition based on the measured momentum of the impinging 
track. Because we are interested in the cumulative distribution averaged over many events 
and because each event contains many tracks, this averaged correction results in a 
negligible contribution to the uncertainty in the measured electromagnetic energy.

We have studied hadron shower spatial and energy distributions in the calorimeter both 
experimentally, using well tracked and identified hadrons in sparse events in STAR, and in 
detailed GEANT simulations.

A library of separate profiles for pions, kaons, protons and antiprotons was obtained from 
GEANT simulations of detector response in the STAR environment (GSTAR). The input events 
had a uniform momentum distribution in the range $0 < p < 10$ GeV/$c$ and an emission 
vertex limited by $|z_{vertex}| < 20$ cm. The constraint on the longitudinal coordinate of 
the vertex insures that the trajectory of particles will extrapolate through only one tower 
of the EMC.  Because the EMC is a projective detector, this constraint on the extrapolated 
track is strongly related to the vertex constraint. We projected the simulated tracks on 
the EMC using a helix model for the particle trajectory in a magnetic field and obtained 
the energy distributions and the corresponding mean values as a function of the momentum, 
the pseudorapidity of the EMC towers and the distance of the incident hit point to the 
center of the tower ($d$). The distributions were binned in intervals of $\Delta\eta=0.2$. 
For all particles, the total mean deposited hadronic energy in a particular tower increases 
approximately linearly with the momentum, shows very little dependence on pseudorapidity 
and decreases with increasing distance from the hit point to the  center of the tower. 
Experimental hadronic shower profiles were obtained from Au+Au minimum bias data by 
projecting tracks on the EMC, accepting only those that were isolated in a 5x5 tower patch 
to ensure that the energy in the towers was from only one particle, and calculating the 
energy distributions and mean values.  Profiles for all particles, except electrons and 
positrons, for both positive and negative tracks were recorded with good statistics up to 
momentum $p = 2.0$ GeV/$c$. 

\begin{figure}
 \includegraphics[width=0.45\textwidth]{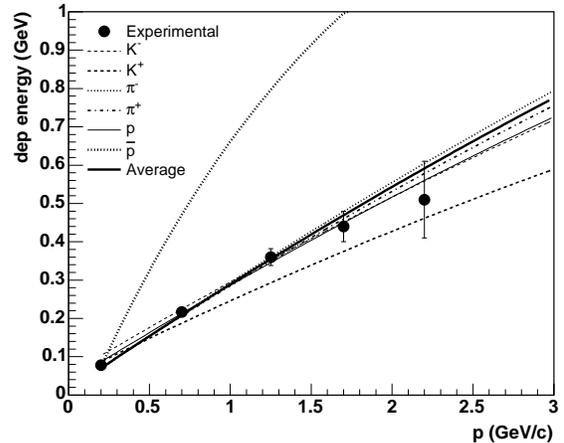}
 \caption{Mean values from GEANT simulations of the energy deposited in the EMC by various 
hadronic species as a function of momentum.}
 \label{figHadDepMasses}
\end{figure}

\begin{figure}
 \includegraphics[width=0.45\textwidth]{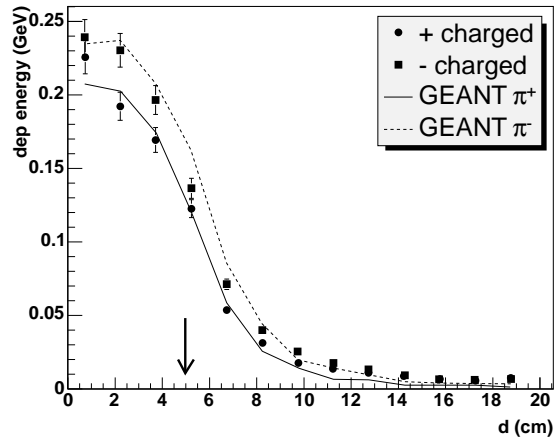}
 \caption{Spatial profiles of energy deposition in the EMC as a function of distance ($d$) 
from the hit point to the center of the tower for $\pi^+$ and $\pi^-$ from simulations and 
for positive and  negative hadrons from data.  The arrow indicates the distance 
corresponding to the border of a tower in $0 < \eta < 0.2$. An overall agreement between  
the shapes of the profiles is observed, with a small normalization difference (see text).}
 \label{figHadDepExpSim}
\end{figure}

In Fig. \ref{figHadDepMasses} we present the deposited energy for different particles from 
GEANT simulations as a function of momentum, for a fixed pseudorapidity and distance to the 
center of the tower. An average curve, based on the relative yield of the different 
particles is also presented. Small differences are observed for most particles, except for 
the antiproton for which the additional annihilation energy is apparent. The solid points 
are deposited energy obtained from experimental data for charged hadrons. The experimental 
profiles for charged hadrons agree quite well with the averaged profile. Because of the 
limited statistics, it was not possible to obtain the experimental profiles for identified 
hadrons. In Fig. \ref{figHadDepExpSim} we present the simulated profiles for $\pi^+$ and 
$\pi^-$ and the experimental profiles for all positively and negatively charged tracks in 
the momentum range $0.5 < p < 1.0$ GeV/$c$, as a function of the distance to the center of 
a tower.  The experimental profiles are well described by the simulation, except for a 
normalization factor on the order of 20\% for $0 < p < 0.5$ GeV/$c$ and 5\% for $p > 0.5$ 
GeV/$c$ , as seen in Fig. \ref{figHadDepExpSim}. After renormalization all experimental 
profiles up to momentum $p = 2.0$ GeV/$c$  are in good agreement with simulation and we 
therefore use the renormalized simulated profiles to allow smooth interpolation in the data 
analysis and for extrapolation to allow corrections for higher momentum tracks. However, 
since the interval $p < 2.0$ GeV/$c$ contains 98\% of all tracks, the magnitude of this 
extrapolation is small for the \ET\ measurement.  

\subsection{$E_T^{em}$ measurement}

The electromagnetic transverse energy is defined as:

\begin{equation}
\label{eqEmDef}
 E_T^{em}=\sum_{towers}E_{tower}^{em} \sin(\theta_{tower})
\end{equation}
where $E_{tower}^{em}$ is the electromagnetic energy measured in an EMC tower and 
$\theta_{tower}$ is the polar angle of the center of the tower relative to the beam axis 
and the collision vertex position. Experimentally, \emET\ is given by: 

\begin{equation}
 \label{eqEmEt}
 E_T^{em}=\frac{1}{f_{acc}}\sum_{towers}(E_{tower}-\Delta E_{tower}^{had}) 
\sin(\theta_{tower})
\end{equation}

\noindent
The sum over EMC towers corresponds to $0 < \eta < 1$ and $\Delta\phi = 60^\circ$. 
$f_{acc}=\Delta\phi/2\pi$ is the correction for the acceptance, $E_{tower}$ is the energy 
measured by an EMC tower, $\Delta E_{tower}^{had}$ is the total correction for each tower 
to exclude the contribution from hadrons . The $\Delta E_{tower}^{had}$ correction is given 
by:

\begin{equation}
 \label{eqHadDep}
 \Delta
 E_{tower}^{had}=\frac{1}{f_{neutral}} \sum_{tracks}\frac{f_{elec}(p_T)}{\mathrm{eff}(p_T)} 
\Delta
 E(p,\eta,d)
\end{equation}
where $\Delta E(p,\eta,d)$ is the energy deposited by a track projected on an EMC tower as 
a function of its momentum $p$, pseudorapidity $\eta$ and distance $d$ to the center of the 
tower from the track hit point. $f_{elec}(p_T)$ is a correction to exclude electrons that 
are misidentified as hadrons and, therefore, should not be added to $\Delta 
E_{tower}^{had}$. This correction was estimated using the same procedure described in the 
previous section to exclude real electrons from the $E_{T}^{had}$ measurement. 
$\mathrm{eff}(p_T)$ is the track efficiency, also discussed previously and $f_{neutral}$ is 
the correction to exclude the long-lived neutral hadron contribution. As in the case for 
\hadET, $f_{neutral} = \Delta E^{charged}_{tower}/(\Delta E^{charged}_{tower}+ \Delta 
E^{neutral}_{tower})$ was estimated from the published STAR data at 130 GeV 
\cite{refSTARpBar, refSTARpBarpErr, refSTARNegChHad, refSTARKaon}. In this case, $\Delta 
E^{neutral}_{tower}$  is defined as the  energy deposited by all long lived neutral 
hadrons. The correction factor is $f_{neutral} = 0.86 \pm 0.03$.

The systematic uncertainty due to the track efficiency correction, as previously discussed, 
is 4\%. The hadronic correction for charged tracks, $\Delta E(p,\eta,d)$, is based 
primarily on measured hadronic shower profiles with GEANT simulations used for 
interpolation between measurements and extrapolation beyond $p = 2$ GeV/$c$. The systematic 
uncertainty for this correction to \emET\ is estimated from the observed uncertainties in 
the calculation of the hadronic profile at points in the shower library where full 
measurements were made. A 5\% systematic uncertainty is consistent with the comparison of 
the measured and calculated shower profiles after normalization. Different from the 
hadronic component of transverse energy, there is no correction for $p_{T}$ cutoff in the 
hadronic background subtraction in the electromagentic energy. Such low $p_{T}$ tracks will 
not reach the calorimeter because of the strength of the magnetic field and, therefore, 
will not deposit energy in the detector.

As discussed earlier, the systematic uncertainty due to calibration of the detector is of 
the order of 2\% and clearly this uncertainty contributes directly to the uncertainty in 
\emET. The systematic uncertainty due to the electron background track correction is 
negligible ($ < 0.5\%$). 

The cumulative effect of all uncertainties discussed in this section, which are assumed to 
be uncorrelated, is an overall systematic uncertainty estimate for \emET\ of 8.0\%. All 
corrections and the corresponding systematic uncertainties are summarized in Table 
\ref{tableEmCor}.

\begin{table}[h]
 \caption{Corrections and systematic uncertainties for \emET\ for the 5\% most central 
collisions. The quadrature sum of all systematic uncertainties, including the hadronic 
shower profiles subtraction ($\Delta E(p,\eta,d)$) not shown in the table, results in a 
total systematic uncertainty of 8\%. The upper part of the table shows the global 
correction and the bottom part shows track wise corrections. In this case, the correction 
values for $p_T=0.25$ Gev/$c$ and $1.0$ GeV/$c$ are shown.}
\begin{ruledtabular}
 \begin{tabular}{cc}
 Description & Correction \\
 \hline
 $f_{neutral}$         & 0.86 $\pm$    0.03 \\
 \hline
 $f_{elec}(p_T)$       & 0.96 $\pm$ $< 0.005$ (0.25 GeV/$c$) \\
                       & 1.00 $\pm$ $< 0.005$ (1 GeV/$c$) \\
 $\mathrm{eff}(p_T)$   & 0.70 $\pm$    0.04   (0.25 GeV/$c$) \\
                       & 0.80 $\pm$    0.04   (1 GeV/$c$) \\
 \end{tabular}
 \end{ruledtabular}
\label{tableEmCor}
\end{table}

\begin{figure}[h]
 \includegraphics[width=0.45\textwidth]{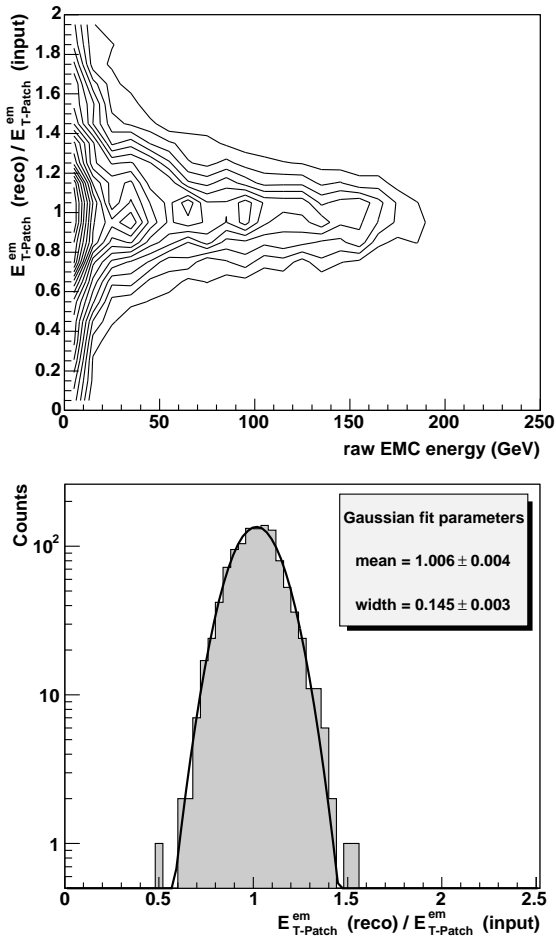}
 \caption{Upper panel - Event-by-event ratio of the reconstructed electromagnetic energy 
and the input from the event generator as a function of the raw energy measured by the EMC. 
At 150 GeV, count numbers vary from 10 to 40 counts from the outer to the inner contour 
lines in steps of $\sim$ 10 counts. Lower panel - The same ratio distribution for the most 
central events. The solid line is a gaussian fit.}
 \label{figEmEtDistribEbyE}
\end{figure}

In order to evaluate the hadronic background subtraction procedure and estimate the 
event-by-event resolution of the reconstructed electromagnetic energy we have performed 
simulations in which we compare the reconstructed \emET\ energy and the input from the 
event generator (HIJING). Fig. \ref{figEmEtDistribEbyE} (upper panel) shows the ratio, 
event by event, of the reconstructed to the input electromagnetic transverse energy as a 
function of the raw energy measured by the calorimeter in the same acceptance used in this 
analysis. The smaller the raw EMC energy, the larger the impact parameter of the collision. 
The reconstructed energy, on average, is the same as the input from the event generator. 
Edge effects due to the limited acceptance of the detector were also studied and the effect 
on the reconstructed values, on average, are negligible. The event-by-event resolution, 
however, improves as the event becomes more central. Fig \ref{figEmEtDistribEbyE} (lower 
panel) shows the ratio distribution for the most central events. The solid line is a 
gaussian fit, from which we estimate that the event-by-event resolution  of the 
reconstructed electromagnetic energy to be 14.5\% for central events. The main factors that 
determine this resolution are the hadronic energy subtraction and the corrections for track 
efficiency and long-lived neutral hadrons. The effect on the global measurement due to the 
tower energy resolution, considering the EMC patch available, was estimated to be 0.5\% and 
that due to calibration fluctuations is 0.5\%. The fluctuations due to the hadronic 
background subtraction procedure alone were estimated to be 12\%, strongly dependent on the 
number of tracks used to correct the energy (for larger acceptances this resolution 
improves). The final \emET\ distribution is also corrected for vertex reconstruction 
efficiency.

\section{Total Transverse Energy $E_{T}$}

\begin{figure}[t]
 \includegraphics[width=0.45\textwidth]{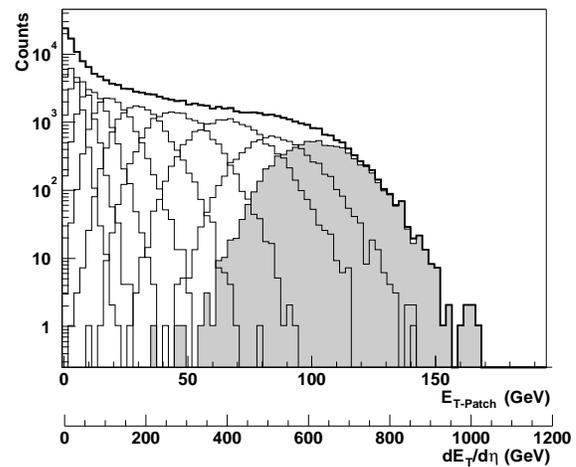}
 \caption{Total Transverse Energy for $0<\eta<1$. The minimum bias distribution is  
presented as well as the distributions for the different  centrality bins (see table 
\ref{Table01}). The shaded area corresponds to the 5\% most central bin. The main axis 
scale corresponds to the \ET\ measured in the detector acceptance and the bottom axis is 
corrected to represent the  extrapolation to full azimuthal acceptance.}
 \label{figTotEtDistrib}
\end{figure}

The sum of \hadET\ and \emET\ is the total transverse energy \ET\ of the events. In Fig. 
\ref{figTotEtDistrib} we present the \ET\ distribution for minimum bias events, corrected 
for vertex reconstruction efficiency mainly in the low \ET\ region. The scale of the upper 
horizontal axis corresponds to the \ET\ measurement for the actual acceptance of $0 < \eta 
< 1$ and $\Delta\phi = 60^\circ$. The bottom axis is scaled to correspond to the \ET\ for 
full azimuthal coverage. 

In Fig. \ref{figTotEtDistrib} we also present the \ET\ distributions for different 
centrality bins defined by the percentages of the total cross section, selected on charged 
multiplicity with $|\eta|<0.5$. The centrality bin defined as 0-5\% (shaded area in Fig. 
\ref{figTotEtDistrib}) corresponds to the most central collisions amounting to 5\% of the 
total cross section. The data for these centrality ranges are given in Table \ref{Table01}. 
The centrality bins are determined by the uncorrected number of charged tracks with 
$|\eta|<0.5$ and number of fit points larger than 10. The phase space overlap between the 
\ET\ and centrality measurements is small so that there is negligible correlation between 
them beyond that due to the collision geometry.

At the low energy edge, the distribution exhibits a peak, corresponding to the most 
peripheral collisions. For the largest values of \ET\, the shape of the distribution is 
determined largely by statistical fluctuations and depends greatly on the experimental 
acceptance \cite{refAcc}. For larger acceptances, the decrease with increasing \ET\ is very 
sharp. For this measurement, the fall off of the distribution at large \ET\ is strongly 
dominated by the limited acceptance which, at this point, obscures any possible physics 
fluctuation. Combining the two contributions (hadronic and electromagnetic energies) to the 
total transverse energy and properly taking into account the correlated uncertainties, we 
estimate a combined systematic uncertainty in $E_T $ of 7\% and an event-by-event 
resolution of 17\%. We obtained for the 5\% most central collisions $\langle 
dE_{T}/d\eta|_{\eta=0.5}\rangle = \langle E_T \rangle _{5\%} =621 \pm 1(stat)\pm 43(syst)$ 
GeV, scaled for full azimuthal acceptance and one unit of pseudorapidity.

\begin{table*}[tbp]
\caption{$E_T$ and $E_T^{em}$ as a function of the centrality of the collision. Global 
normalization uncertainties are indicated in the header of the table. All uncertainties are 
systematic. Statistical errors are negligible.}
\label{Table01}
\begin{ruledtabular}
\begin{tabular}{cccccccc}
Centrality (\%) & $N_{part}$ & $N_{bin}$ & $E_{T}$(GeV) & $E_{T}^{em}$(GeV) & 
$E_{T}/N_{ch}$(GeV) &$E_{T}/0.5N_{part}$(GeV) &  $E_{T}^{em}/E_{T}$ \\
      & & &    $\pm 4.3\%$ & $\pm 4.8\%$ & $\pm 5.1\%$  & $\pm 4.3\%$  & $\pm 3.4\%$ \\ 
\hline
$70-80$ & $ 14\pm  4$ & $ 12 \pm  4$ & $17.1 \pm 0.9$ & $  5.8\pm   0.4$ & $  0.69\pm 0.07$ 
& $   2.4\pm  0.6$  & $  0.342\pm 0.031$ \\
$60-70$ & $ 27\pm  5$ & $ 29 \pm  8$ & $37.6 \pm 2.0$ & $ 13.4\pm   0.9$ & $  0.75\pm 0.07$ 
& $   2.8\pm  0.5$  & $  0.357\pm 0.022$ \\
$50-60$ & $ 47\pm  8$ & $ 64 \pm 14$ & $70   \pm   4$ & $ 25.9\pm   1.7$ & $  0.79\pm 0.06$ 
& $   3.0\pm  0.5$  & $  0.369\pm 0.020$ \\
$40-50$ & $ 76\pm  8$ & $123 \pm 22$ & $118  \pm   6$ & $ 43  \pm   3$   & $  0.82\pm 0.06$ 
& $   3.1\pm  0.4$  & $  0.364\pm 0.020$ \\
$30-40$ & $115\pm  9$ & $220 \pm 30$ & $187  \pm  10$ & $ 68  \pm   4$   & $  0.85\pm 0.06$ 
& $   3.2\pm  0.3$  & $  0.362\pm 0.019$ \\
$20-30$ & $166\pm  9$ & $368 \pm 41$ & $279  \pm  15$ & $100  \pm   6$   & $  0.86\pm 0.06$ 
& $   3.31\pm 0.25$ & $  0.357\pm 0.019$ \\
$10-20$ & $234\pm  8$ & $591 \pm 52$ & $402  \pm  21$ & $143  \pm   9$   & $  0.86\pm 0.06$ 
& $   3.40\pm 0.22$ & $  0.356\pm 0.019$ \\
$5-10$  & $299\pm  7$ & $828 \pm 64$ & $515  \pm  28$ & $181  \pm  12$   & $  0.86\pm 0.06$ 
& $   3.43\pm 0.20$ & $  0.351\pm 0.019$ \\
$0-5$   & $352\pm  3$ & $1051\pm 72$ & $620  \pm  33$ & $216  \pm  14$   & $  0.86\pm 0.06$ 
& $   3.51\pm 0.19$ & $  0.348\pm 0.019$ \\
\end{tabular}
\end{ruledtabular}
\end{table*}

\begin{figure} [t]
 \includegraphics[width=0.45\textwidth]{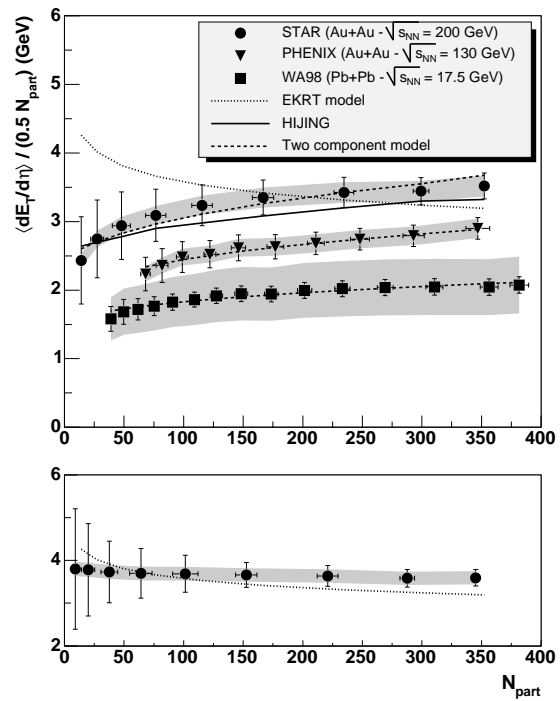}
 \caption{$\langle dE_{T}/d\eta|_{\eta=0.5}\rangle$ per  $N_{part}$ pair vs. $N_{part}$. 
Upper panel: $N_{part}$ is obtained from Monte Carlo Glauber calculations. The lines show 
calculations using the HIJING model \cite{refHIJING}(solid) , the EKRT saturation model 
\cite{refEskola} (dotted, Eq. \ref{eqEskola}) and the two component fit (dashed, see text). 
Results from WA98 \cite{refWA98} and PHENIX \cite{refPHENIX} are also  shown. The grey 
bands correspond to overall systematic  uncertainties, independent of $N_{part}$. Error 
bars are the quadrature sum of the errors on the measurements and the uncertainties on 
$N_{part}$ calculation.  Lower panel: the same data are shown as in the upper panel but 
using and Optical Glauber model calculation for \Npart. The line shows the same result from 
EKRT model calculation.}
 \label{figdEtdEtaPerPairNpart}
\end{figure}

The upper panel of Fig. \ref{figdEtdEtaPerPairNpart} shows $\langle dE_T/d\eta \rangle$ per 
participant pair $N_{part}/2$ as a function of \Npart\ (obtained using Monte Carlo Glauber 
calculations \cite{refNpartNbin}). Data from Au+Au collisions at $\sqrt{s_{NN}}$= 200 GeV 
from this analysis are shown together with similar measurements from Pb+Pb collisions at 
$\sqrt{s_{NN}}$=17.2 GeV from WA98 \cite{refWA98} and Au+Au collisions at 130 GeV from 
PHENIX \cite{refPHENIX}. These comparison measurements are at $\eta$=0, whereas the 
measurement reported here is at $0<\eta<1$. The grey bands for all three datasets show the 
overall systematic uncertainty of the data independent of \Npart, while the error bars show 
the quadratic sum of the statistical errors, which are typically negligible, and the 
systematic uncertainties in \ET\ and \Npart\ \cite{refNpartNbin}, with the latter 
dominating at low \Npart.

A model based on final state gluon saturation (EKRT \cite{refEskola}) predicts a decrease 
in more central nuclear collisions for both the charged particle multiplicity per 
participant and \ET. Hydrodynamic work during expansion may reduce the observed $E_T$ 
relative to the initially generated $E_T$,perhaps by a factor $\sim3$ at RHIC energies, 
\cite{refEskola}, though this effect will be offset somewhat by the buildup of transverse 
radial flow \cite{refEskola,refKolb}. The dependence of observed \ET\ in $\sqrt{s}$ and 
system size $A$ in the EKRT model is:

\begin{equation}
\label{eqEskola}
E_T^{b=0} = 0.43 A^{0.92}(\sqrt{s})^{0.40}(1-0.012 \ln A +0.061 \ln\sqrt{s})
\end{equation}

\noindent
The centrality dependence can be approximated by replacing $A$ by \Npart/2 
\cite{refEskola}, shown by the dotted line in Fig \ref{figdEtdEtaPerPairNpart}. The upper 
panel shows comparison to measured \dETdeta\ per participant pair, incorporating a Monte 
Carlo Glauber calculation for \Npart. The EKRT model is seen not to agree with the data in 
this panel, missing significantly both the centrality dependence and the normalization for 
central collisions. A similar comparison is made in the lower panel of Fig. 
\ref{figdEtdEtaPerPairNpart}, which differs from the upper panel only in the use of an 
Optical Glauber calculation for \Npart\ \cite{refNpartNbin}. The centrality dependence of 
the data in this case is reproduced well by the model, though $\sim15\%$ disagreement in 
normalization for central collisions remains. More precise comparison of the system size 
dependence of \ET predicted by EKRT model to RHIC data requires either further refinement 
of the Glauber model calculations or measurements for central collisions with varying mass 
$A$.

The HIJING model predicts an increase in \avdETdeta/(0.5\Npart), as shown in Fig 8 upper 
panel. HIJING incorporates hard processes via the generation of multiple minijets together 
with soft production via string fragmentation. Effects of the nuclear geometry in HIJING 
are calculated using the Monte Carlo Glauber approach. Agreement of HIJING with the data is 
seen to be good in the upper panel.

We also study a simple two-component approach where $dE_T/d\eta = A  N_{part} + B  
N_{bin}$. Using this model it is possible to estimate the fraction of hard collisions in 
the \ET\ production. In this case, a simple fit function: 

\begin{equation}
\label{eqFit}
dE_T/d\eta/(0.5 N_{part})= 2A [1+(B/A) (N_{bin}/N_{part})] 
\end{equation}
is applied to our data at 200 GeV and the published PHENIX and WA98 results including 
points with number of participants larger than 100. The results from the fits are shown in 
Table \ref{tableFit}. The simple scaling ansatz does a good job describing the overall 
shape of the \Npart\ dependence at all energies. In this picture, the ratio 
$(B/A)(N_{bin}/N_{part})$ estimates the fraction of the transverse energy that scales like 
hard processes. As seen in the third column in Table \ref{tableFit} this ratio for the most 
central events is constant within errors despite the expectation that the cross section for 
hard processes grows by a large factor from 17 to 200 GeV. 

\begin{table} [t]
 \caption{Two component model fit results of $dE_T/d\eta = A N_{part} + B  N_{bin}$. The 
uncertainties in the fit parameters include both the data and the $N_{part}$($N_{bin}$) 
uncertainties.} 
\begin{ruledtabular}
 \begin{tabular}{cccc}
  & $A$ (Gev) & $B/A$ & $(B/A) (N_{bin}/N_{part})$\\
 \hline
 STAR    & $1.21 \pm 0.21$ & $0.17 \pm 0.09$ & $0.55 \pm 0.14$     \\
 PHENIX  & $0.83 \pm 0.18$ & $0.27 \pm 0.15$ & $0.71 \pm 0.32$     \\
 WA98    & $0.66 \pm 0.16$ & $0.28 \pm 0.11$ & $0.59 \pm 0.23$     \\
 \end{tabular}
\end{ruledtabular}
\label{tableFit}
\end{table}

\begin{figure}[t]
 \includegraphics[width=0.45\textwidth]{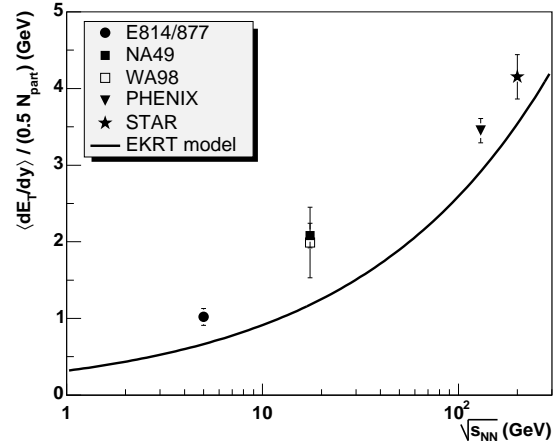}
 \caption{$dE_T/dy$ (see text for details) per $N_{part}$ pair vs $\sqrt{s_{NN}}$ for 
central events. In this  figure $dE_T/dy/(0.5 N_{part})$ is seen to grow logarithmicaly 
with $\sqrt{s_{NN}}$. The error bar in the STAR point represents the total systematic 
uncertainty. The solid line is a EKRT model prediction \cite{refEskola}, corrected for 
$d\eta/dy$, for central Au+Au collisions.}
 \label{figdEtdEtaPerPairNpartXsqrtS}
\end{figure}

We observe an overall increase in the transverse energy of $(24 \pm 7)\%$ at 200 GeV 
relative to 130 GeV.  In Fig. \ref{figdEtdEtaPerPairNpartXsqrtS} we present our result for 
$dE_T/dy$ per participant pair for 
central collisions, together with results from other experiments at various collision 
energies from AGS to RHIC \cite{refBarrete2, refWA98, refPHENIX, refNA49}.  For the 
purposes of this comparison, we calculated $dE_T/dy$ from $dE_T/d\eta$ for our measurements 
using a factor of 1.18 obtained from HIJING simulations to convert from $\eta$ to $y$ 
phase-space.  Our result is consistent with an overall logarithmic growth of $dE_T/dy/(0.5 
N_{part})$ with $\sqrt{s_{NN}}$. The solid line is the prediction using the EKRT model 
\cite{refEskola} for central Au+Au collisions.  As one can see, EKRT model underestimates 
the final transverse energy by $\sim 15\%$.

We have also estimated the spatial energy density produced in the collision using $\langle 
E_T \rangle _{5\%}$ reported above, converted from pseudorapidity to rapidity density using 
the factor of 1.18 discussed above. Based on a scaling solution to the relativistic 
hydrodynamic equations, Bjorken \cite{refBjorken} estimated the  spatial energy density of 
the system in terms of the primordial transverse energy rapidity density $dE_{T}/dy$, the 
transverse system size, $R$, and a formation time  $\tau_{0}$. 

\begin{equation}
\label{eqBjorken}
 \varepsilon_{\textit{Bj}}=\frac{dE_T}{dy}\frac{1}{\tau_0\pi R^2}
\end{equation}

We assumed $\tau_0 = 1$ fm/\textit {c} which is the usual value taken in many analysis at 
SPS energies. For Au+Au at $\sqrt{s_{NN}} = 200$ GeV we obtained 
$\varepsilon_{\textit{Bj}}=4.9 \pm 0.3$ GeV/fm$^3$. The uncertainty includes only the 
uncertainty on $\langle dE_{T}/d\eta \rangle$. This energy density is significantly in 
excess of the energy density $\sim1$ GeV/fm$^3$ predicted by Lattice QCD for the transition 
to a deconfined Quark Gluon Plasma \cite{refKarsch}. The estimate is based however upon the 
assumption that local equilibrium has been achieved at $\tau\sim1$ fm/c and that the system 
then expands hydrodynamically. Comparison of other RHIC data, in particular elliptic flow, 
to hydrodynamic calculations \cite{refSTARv2,refKolb2} indicates that this picture may 
indeed be valid.

In order to understand the systematic growth in transverse energy  with collision energy 
shown in Fig. \ref{figdEtdEtaPerPairNpartXsqrtS} we investigate the centrality dependence 
of \avdETdeta/\avdNchdeta, the scaling of transverse energy relative to the number of 
charged particles produced in the collision. The centrality dependence of this ratio may 
indicate effects of hydrodynamic flow \cite{refKolb}: if the expansion is isentropic then 
\dNchdeta\ will remain constant, whereas \dETdeta\ will decrease due to the performance of 
longitudinal work.

\begin{figure}[t]
 \includegraphics[width=0.43\textwidth]{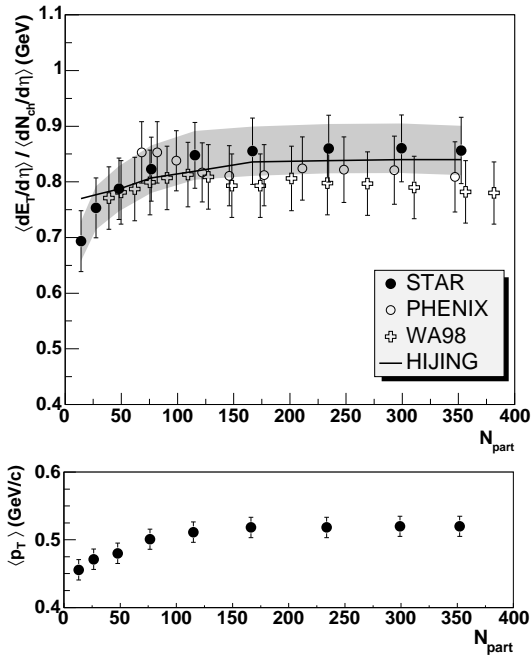}
 \caption{Upper panel - $\langle dE_T/d\eta \rangle / \langle dN_{ch}/d\eta \rangle$ vs  
$N_{part}$. Predictions from HIJING simulations for Au+Au at 200 GeV are presented. Results 
from WA98 \cite{refWA98} and PHENIX \cite{refPHENIX} are also  shown. The grey band 
corresponds to an overall normalization uncertainty for the STAR measurement. Bottom panel 
- Charged hadrons mean transverse momentum as a function of $N_{part}$ \cite{refMeanPt200}. 
}
 \label{figdEtdEtaTodNchdEtaAndMeanPt}
\end{figure} 
 
\begin{figure}[t] 
 \includegraphics[width=0.45\textwidth]{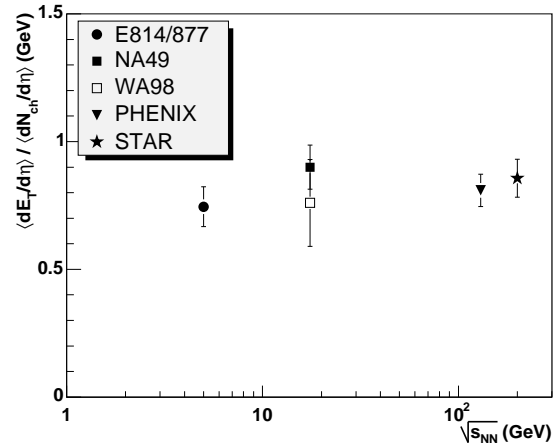}
 \caption{$\langle dE_T/d\eta \rangle / \langle dN_{ch}/d\eta \rangle$ vs $\sqrt{s_{NN}}$ 
for central events. The error bar in the STAR point corresponds to the  systematic 
uncertaintiy. A  constant value of $\sim 800$ MeV per charged particle, within errors,  
characterizes transverse energy production over this full energy  range.}
 \label{figdEtTodNchXsqrtS}
\end{figure}

Fig. \ref{figdEtdEtaTodNchdEtaAndMeanPt}, upper panel, shows the centrality dependence of 
\avdETdeta/\avdNchdeta\ from STAR measurements at $\sqrt{s_{NN}}=200$ GeV, compared to 
similar measurements at 17 and 130 GeV. Data at all energies fall on a common curve within 
uncertainties, with modest increase from the most peripheral collisions to \Npart=100, 
reaching a roughly constant value of \avdETdeta/\avdNchdeta. Fig. 
\ref{figdEtdEtaTodNchdEtaAndMeanPt}, lower panel, shows the $\langle p_T \rangle$ for 200 
GeV Au+Au collisions measured by STAR \cite{refMeanPt200}, showing a dependence on 
centrality similar to that of the transverse energy per charged particle: modest increase 
with \Npart\ for $\Npart<100$, with constant value for more central collisions. The 
systematic behavior of $E_T$, multiplicity and $\langle p_T \rangle$ are similar, 
indicating that the growth of $E_T$ is due to increased particle production. Quantitative 
comparison of theoretical models of particle production with the measured centrality 
dependences of \avdETdeta/\avdNchdeta\ and $\langle p_T \rangle$ of charged particles will 
constrain the profile of initial energy deposition and the role of hydrodynamic work during 
the expansion.

In Fig. \ref{figdEtTodNchXsqrtS} we show, for central collisions, that this constant 
transverse energy production  per charged particle is observed down to and including AGS 
measurements at $\sqrt{s_{NN}} = 5$ GeV. A single  value of $\sim 800$ MeV per charged 
particle or at most a slow logarithmic increase amounting to $< 10$\%  characterizes all 
measurements within errors over a range in which the \ET\ per participant grows by a factor 
of four. HIJING predicts that \ET\ per charged particle should increase from SPS to RHIC 
energies due to the  enhancement of minijet production at RHIC. However, the predicted 
increase is rather small and the systematic  uncertainties on the measurement do not 
provide enough precision to significantly test this assumption.
 
\begin{figure} [t]
 \includegraphics[width=0.45\textwidth]{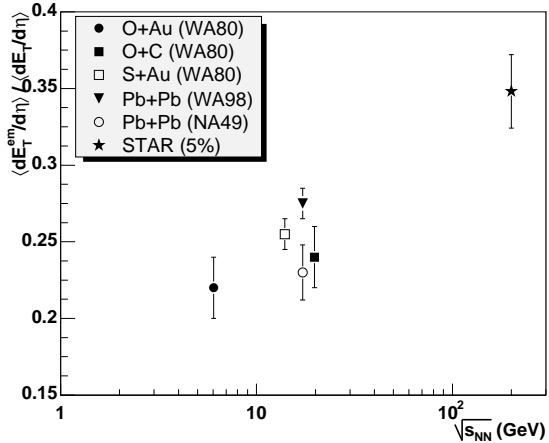}
 \caption{Energy dependence of the electromagnetic fraction of  the transverse energy for a 
number of systems spanning SPS to RHIC energy for central events.}
 \label{figEmEtToTotEtXsqrtS} 
\end{figure}

The procedures adopted in this analysis permit an independent measurement of the 
electromagnetic and hadronic  transverse energy. This allows additional exploration of the 
collision dynamics and particle  production. In Fig. \ref{figEmEtToTotEtXsqrtS} we show the 
ratio of the electromagentic to the total energy for the most central events as a function 
of the energy from lower SPS energies \cite{refWA982, refWA80} to our  results at full RHIC 
energy. The observed electromagnetic fraction of the total transverse energy will be 
strongly influenced by the baryon to meson ratio. At very high energy it is expected that 
virtually all the \ET\ will be carried by mesons and the fraction should approximate $1/3$,  
whereas at low energy, baryon dominance of the transverse energy will result in a much 
smaller electromagnetic  fraction.
 
\begin{figure}[h]
 \includegraphics[width=0.45\textwidth]{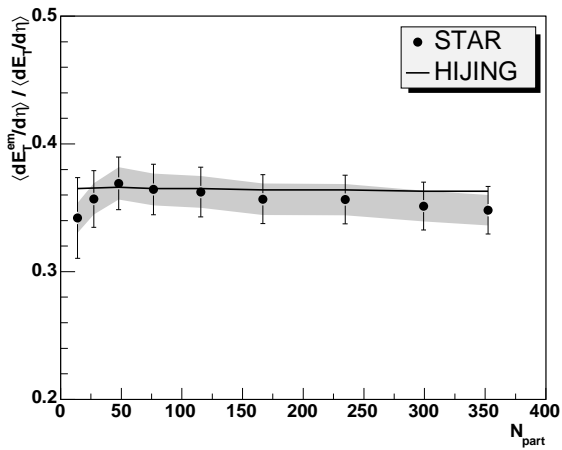}
 \caption{Participant number dependence of the electromagnetic  fraction of the total 
transverse energy.  The results are  consistent with HIJING within errors  over the full 
centrality range.}
\label{figEmEtToTotEtXNpart}
\end{figure}

While the energy dependence seen in Fig. \ref{figEmEtToTotEtXsqrtS} is presumably dominated 
by the total meson content of the final state, the centrality dependence may provide 
additional detail about the reaction mechanisms. The centrality dependence of the 
electromagnetic fraction of our total measured energy is shown in Fig. 
\ref{figEmEtToTotEtXNpart}. An excess photon yield may result from the formation of a 
long-lived deconfined phase, as suggested in Ref. \cite{refDIRPH}. The predictions from 
HIJING simulations are also presented. We observe no significant dependence of the 
electromagnetic fraction with the collision centrality.

\section{Summary}

We have reported the measurement of transverse energy $E_T$ within $0<\eta<1$, for 
centrality-selected Au+Au collisions at $\sqrt{s_{NN}}=200$ GeV. For the 5\% most central 
events we measured $\langle E_{T} \rangle _{5\%} = 621 \pm 1(stat)\pm 43(syst)$ GeV, 
corresponding to an increase of (24$\pm$7)\% with respect to measurements at 130 GeV Au+Au 
collisions at RHIC \cite{refPHENIX}. 

We investigated the energy scaling with the number of participant nucleons and with the 
number of charged particles produced in the  collision. We obtained, for the 5\% most 
central events, $dE_{T}/d\eta/(0.5N_{part}) =
3.51 \pm 0.24$ GeV and $\langle dE_{T}/d\eta \rangle/\langle dN_{ch}/d\eta \rangle = 860 
\pm 70$ MeV, respectively. We also compared the results of this work with measurements from 
AGS and SPS energies. It was found that the increase in the \ET\ production from AGS up to 
RHIC energies comes mostly from the increase in the particle production. A final state 
gluon saturation model (EKRT), HIJING and a simple two component (hard/soft) model were 
compared to the data. Although the EKRT model predicts a different centrality behavior of 
energy production the uncertainties in the \Npart\ determination does not allow us to 
discard this model. The simple two component ansatz suggests that, despite the large 
uncertainties, the fraction of energy arising from hard processes which is still visible in 
the final state does not increase significantly from SPS to RHIC energies. 

Other measurements at RHIC and comparison to theoretical calculations suggest that a dense, 
equilibrated system has been generated in the collision and that it expands as an ideal 
hydrodynamic fluid. The good agreement between hydrodynamic calculations and measurements 
of particle-identified inclusive spectra and elliptic flow \cite{refSTARv2} are consistent 
with the onset of hydrodynamic evolution at a time $\tau_0 < 1$ fm/c after the collision 
\cite{refKolb2}. The strong suppression phenomena observed for high \pT\ hadrons 
\cite{refSTARHighpt1,refSTARHighpt2,refSTARHighpt3} suggest that the system early in its 
evolution is extremely dense. Estimates based on these measurements yield an initial energy 
density in the vicinity of 50-100 times cold nuclear matter density. Within the framework 
of boost-invariant scaling hydrodynamics \cite{refBjorken}, from the \ET\ measurement 
presented here we estimate an initial energy density of about 5 GeV/fm$^3$. This should be 
understood as a lower bound \cite{refMatsui,refDumitru}, due to the strong reduction in the 
observed relative to the initially produced \ET\ from longitudinal hydrodynamic work during 
the expansion. These three quite different approaches produce rough agreement for the 
estimated initial energy density, with a value well in excess of that predicted by Latice 
QCD for the deconfinement phase transition \cite{refKarsch}.

The method used in this analysis permitted an independent measurement of  the 
electromagnetic and hadronic components of the total energy. The electromagnetic fraction 
of the transverse energy for the 5\% most central events obtained in this work is $\langle 
dE_{T}^{em}/d\eta \rangle\//\langle dE_{T}/d\eta \rangle = 0.35 \pm 0.02$, consistent with 
a final state dominated by mesons. Some models \cite{refDIRPH} expect that the formation of 
a long lived deconfined phase in central events may increase the yield of direct photon 
production and, therefore, an increase in the electromagnetic fraction of the transverse 
energy. We, however, observe that the electromagnetic fraction of the transverse energy is 
constant, within errors, as a function of centrality. Measurements with larger acceptances 
would have systematic uncertainties significantly reduced and therefore would be able to 
show smaller effects that can not be observed with the precision of the present 
measurement.

We thank the RHIC Operations Group and RCF at BNL, and the NERSC Center at LBNL for their 
support. This work was supported in part by the HENP Divisions of the Office of Science of 
the U.S. DOE; the U.S. NSF; the BMBF of Germany; IN2P3, RA, RPL, and EMN of France; EPSRC 
of the United Kingdom; FAPESP of Brazil; the Russian Ministry of Science and Technology; 
the Ministry of Education and the NNSFC of China; Grant Agency of the Czech Republic, FOM 
and UU of the Netherlands, DAE, DST, and CSIR of the Government of India; Swiss NSF; and 
the Polish State Committee for Scientific Research.

\end{document}